\documentclass[12pt,a4paper]{article}
\usepackage{epsfig}
\begin{document}
\title{The littlest Higgs model and Higgs boson associated
production with top quark pair at high energy \\
linear $e^{+}e^{-}$ collider}

\author{Chong-Xing Yue, Wei Wang,  Feng Zhang\\
{\small  Department of Physics, Liaoning Normal University, Dalian
116029, China}\thanks{E-mail:cxyue@lnnu.edu.cn}\\}
\date{\today}

\maketitle
\begin{abstract}

In the parameter space allowed by the electroweak precision
measurement data, we consider the contributions of the new
particles predicted by the littlest Higgs($LH$) model to the Higgs
boson associated production with top quark pair in the future high
energy linear $e^{+}e^{-}$ collider($ILC$). We find that the
contributions mainly come from the new gauge bosons $Z_{H}$ and
$B_{H}$. For reasonable values of the free parameters, the
absolute value of the relative correction parameter
$\delta\sigma/\sigma^{SM}$ can be significanly large, which might
be observed in the future $ILC$ experiment with $\sqrt{S}=800GeV$.
\vspace{2cm}

\hspace{-0.5cm}PACS number:12.60.Cn, 14.80.Cp, 14.65.Ha

\end{abstract}

\newpage
\noindent{\bf I. Introduction}

The mechanism of electroweak symmetry breaking($EWSB$) remains the
most prominent mystery in current particle physics despite of the
success of the standard model($SM$) tested by high energy
experimental data. The $SM$ accommodates fermion and weak gauge
boson masses by including a fundamental Higgs scalar $H$, which is
assumed to responsible to break the electroweak symmetry. However,
the $SM$ suffers from problems of triviality, unnaturalness, etc.
Thus, the $SM$ can only be an effective field theory below some
high energy scales. New physics should exist at energy scales
around $TeV$. Studying $EWSB$ mechanisms other than the simple
$SM$ Higgs sector is one of the interesting topics in current
particle physics. Little Higgs models[1,2,3] were recently
proposed as one kind of models of $EWSB$. The key feature of this
kind of models is that the Higgs boson is a pseudo-Goldstone boson
of a global symmetry which is spontaneously broken at some higher
scale $f$ and thus is naturally light. $EWSB$ is induced by a
Coleman-Weinberg potential, which is generated by integrating out
the heavy degrees of freedom. This type of models can be regarded
as one of the important candidates for the new physics beyond the
$SM$.

The precision electroweak measurement data suggest that the Higgs
boson must be relative light and its mass should be roughly in the
range of $114.4GeV\sim 260GeV$ for $m_{t}= 178GeV$ at $95\% C. L.$
[4]. The discovery and study of Higgs boson is one of the most
important goals of present and future high energy collider
experiments. The $LHC$ will make the first exploration of the
$TeV$ energy range, and will be able to discover Higgs boson in
the full mass range, provided it exists[5]. After the discovery of
the Higgs boson at the $LHC$, one of the most pressing tasks is a
proper determination of the properties of this scalar since it is
very important to study the mechanism of $EWSB$ and the generation
of mass. The $LHC$ will be able to finish a few measurement on the
couplings of the Higgs boson to fermions and gauge bosons but the
most precise measurements will be performed in the clean
environment of a future high energy linear $e^{+}e^{-}$ collider,
the International Linear Collider($ILC$)[6].

The top quark, with a mass of the order of the electroweak scale
$m_{t}\approx 178.0\pm4.36GeV$[7], is the heaviest particle yet
discovered. The coupling of Higgs boson to top quark pair, which
is the largest one among the Yukawa couplings, should be detected
at high energy experiments. This coupling should play a key role
in a theory generating fermion masses and is particularly
sensitive to the underlying physics. Thus, studying the
$ht\overline{t}$ Yukawa coupling is of particular interest, which
is helpful to precision test the $SM$ and search for new physics
beyond the $SM$.

The Higgs boson associated production with top quark pair
$t\overline{t}$ at the hadron or lepton colliders plays a very
important role both for discovery and for precision measurements
of the Yukawa coupling $t\overline{t}h$. Such measurements could
help to distinguish the $SM$ Higgs boson from more complex Higgs
sectors and shed light on the details of the generation of fermion
masses[8]. At hadron colliders, such as $LHC$, Higgs boson can be
associated production with $t\overline{t}$ pair through $gg$ and
$q\overline{q}$ sub-process, which has been extensively studied in
Ref.[9]. Besides hadron colliders, the $ILC$ can also produce a
Higgs boson with $t\overline{t}$ pair via the process
$e^{+}e^{-}\rightarrow t\overline{t}h$, as long as the Higgs mass
is not too large, i.e., $M_{h}\sim 114.4GeV\sim 260GeV$[10]. The
process $e^{+}e^{-}\rightarrow t\overline{t}h$ proceeds mainly
through Higgs-boson emission off top quarks, while emission from
intermediate $Z$ bosons plays only a minim role. Thus, it is
suitable to determinate the coupling $g_{t\overline{t}h}$. If the
Higgs boson is light, a precision of around $5\%$ can be reached
at an $ILC$ with the c.m. energy $\sqrt{S}=800GeV$ and the
integrated luminosity $\pounds_{int}\approx 1000fb^{-1}$[6].
Furthermore, by studying the contributions of the new physics to
the process $e^{+}e^{-}\rightarrow t\overline{t}h$, one can obtain
the bounds on the free parameters of the non-standard models[11].
The purpose of this paper is to calculate the corrections of new
particles predicted by the littlest Higgs$(LH)$ model[1] to the
process $e^{+}e^{-}\rightarrow t\overline{t}h$ and see whether the
effects on this process can be observed in the future $ILC$
experiment with $\sqrt{S}=800GeV$.

\noindent{\bf II. $t\overline{t}h$ production cross section in the
$LH$ model}

The $LH$ model is embedded into a non-linear $\sigma$ model with
the coset space of $SU(5)/\\ SO(5)$. At the scale $\Lambda_{s}\sim
4\pi f$, the global $SU(5)$ symmetry is broken into its subgroup
$SO(5)$ via a $VEV$ of order $f$, resulting in 14 Goldstone
bosons. The effective field theory of these Goldstone boson is
parameterized by a non-linear $\sigma$ model with gauged symmetry
$[SU(2)\times U(1)]^{2}$, spontaneously broken down to its
diagonal subgroup $SU(2)\times U(1)$, identified as the $SM$
electroweak gauge group. Four of these Goldstone bosons are eaten
by the broken gauge generators, leaving 10 states that transform
under the $SM$ gauge group as a doublet $H$ and a triplet $\Phi$.
This breaking scenario also gives rise to the four new gauge
bosons $W_{H}^{\pm}$, $B_{H}$ and $Z_{H}$. A new vector-like top
quark $T$ is also needed to cancel the divergences from the top
quark loop. All of these new particles playing together can
successfully cancel off the quadratic divergences of the Higgs
boson mass and may produce characteristic signatures at the
present and future collider experiments[12,13].

\begin{figure}[htb]
\vspace{-4cm}
\begin{center}
\epsfig{file=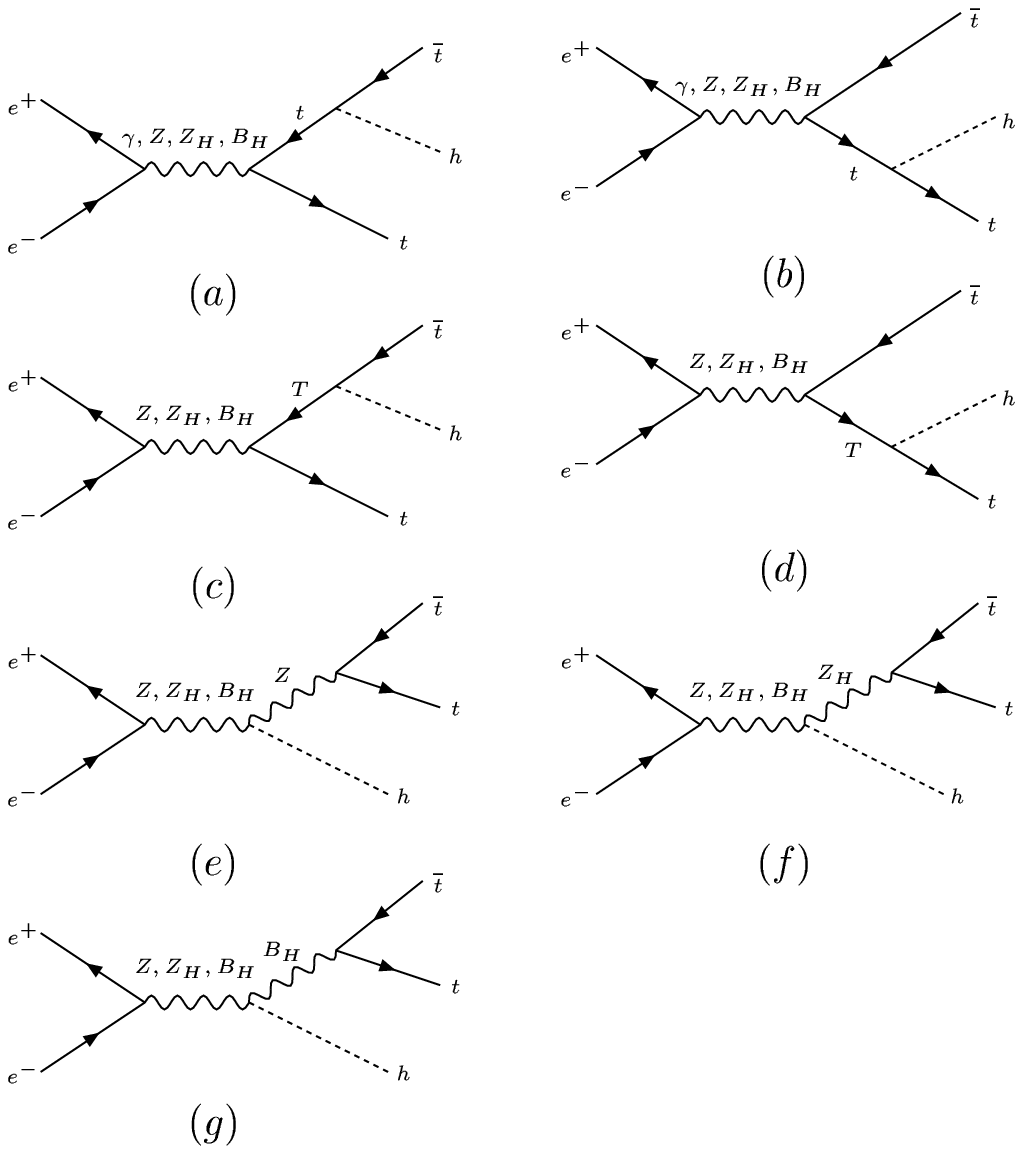,width=600pt,height=780pt} \vspace{-13.75cm}
\hspace{0.5cm} \caption{The Feynman diagrams of the process
$e^{+}e^{-}\rightarrow t\overline{t}h$ in the $LH$ model.}
\label{ee}
\end{center}
\end{figure}

The effective non-linear Lagrangian invariant under the local
gauge group $[SU(2)_{1}\times U(1)_{1}]\times [SU(2)_{2}\times
U(1)_{2}]$, which can be written as [12,14]:
\begin{equation}
{\cal L}_{eff}={\cal L}_{G}+{\cal L}_{F}+{\cal L}_{Y}+{\cal
L}_{\Sigma}-V_{CW},
\end{equation}
where ${\cal L}_{G}$ consists of the pure gauge terms, which can
give the $3-$ and $4-$particle interactions among the $SU(2)$
gauge bosons. The fermion kinetic term ${\cal L}_{F}$ can give the
couplings of the gauge bosons with fermions. The couplings of the
scalars $H$ and $\Phi$ with fermions can be derived from the
Yukawa interaction term ${\cal L}_{Y}$. In the $LH$ model, the
global symmetry prevents the appearance of a Higgs potential at
tree level. The effective Higgs potential, the Coleman-Weinberg
potential $V_{CW}$, is generated at one-loop and higher orders due
to interactions with gauge bosons and fermions, which can induce
to $EWSB$ by driving the Higgs mass squared parameter negative.
${\cal L}_{\Sigma}$ consists of the $\sigma$ model of the $LH$
model.

From the effective non-linear Lagrangian ${\cal L}$, one can
derive the mass and coupling expressions of the gauge bosons,
scalars and the fermions, which have been extensively discussed in
Refs.[12,14]. Using these expressions and the relevant $SM$
Feynman rules, we can calculate the production cross section of
the process $e^{+}e^{-}\rightarrow t\overline{t}h$ in the context
of the $LH$ model.

The relevant Feynman diagrams for the contributions to the
production amplitude of the process $e^{+}e^{-}\rightarrow
t\overline{t}h$ at the order of $\nu^{2}/f^{2}$ in the $LH$ model
are shown in Fig1.$(a)-(g)$, in which $\nu\approx 246GeV$ is the
electroweak scale. Considering the large masses of the  new
particles $T$, $Z_{H}$, $B_{H}$ and the differently Feynman rules,
we can surmise that the total contributions of these new particles
to this process mainly come from Fig.1$(a)$ and $(b)$. We have
confirmed this expectation through explicit calculation. However,
to study the total contributions of the $LH$ model to the
$t\overline{t}h$ production, we will include all contributions in
our numerical calculation. The invariant scattering amplitude of
the process $e(p_{e})+\overline{e}(p_{\overline{e}})\rightarrow
t(p_{t})+\overline{t}(p_{\overline{t}})+h(p_{h})$ can be written
as:
\begin{eqnarray}
M&=&\sum_{v_{i}=\gamma,\ Z,\ Z_{H},\ B_{H}}M_{a}^{v_{i}}+\sum_{v_{i}=\gamma,\
 Z,\ Z_{H},\ B_{H}}M_{b}^{v_{i}}+\sum_{v_{i}=Z,\ Z_{H},\ B_{H}}M_{c}^{v_{i}T}\nonumber\\
&&\hspace{0.5cm}+\sum_{v_{i}=Z,\ Z_{H},\
B_{H}}M_{d}^{v_{i}T}+\sum_{v_{i}=Z,\ Z_{H},\
B_{H}}M_{e}^{v_{i}Z}\\
&&\hspace{0.5cm}+\sum_{v_{i}=Z,\ Z_{H},\
B_{H}}M_{f}^{v_{i}Z_{H}}+\sum_{v_{i}=Z,\ Z_{H},\
B_{H}}M_{g}^{v_{i}B_{H}}\nonumber
\end{eqnarray}
with
\begin{eqnarray}
M_{a}^{v_{i}}&=&\overline{u}_{t}(p_{t})\Lambda_{v_{i}t\overline{t}}^{\mu}
\frac{i[(p_{\overline{t}}+p_{h})\cdot \gamma +
m_{t}]}{(p_{\overline{t}}+p_{h})^{2}
-m_{t}^{2}}\Lambda_{ht\overline{t}}v_{t}
(p_{\overline{t}})\nonumber\\
&&\hspace{1cm}\frac{-ig_{\mu\nu}}{(p_{e}+p_{\overline{e}})^{2}
-M_{v_{i}}^{2}}\overline{v}_{e}(p_{\overline{e}})\Lambda_{v_{i}
e\overline{e}}^{\nu}u_{e}(p_{e});\\
M_{b}^{v_{i}}&=&\overline{u}_{t}(p_{t})\Lambda_{ht\overline{t}}
\frac{i[(p_{t}+p_{h})\cdot \gamma + m_{t}]}{(p_{t}+p_{h})^{2}
-m_{t}^{2}}\Lambda_{v_{i}t\overline{t}}^{\mu}v_{t}
(p_{\overline{t}})\nonumber\\
&&\hspace{1cm}\frac{-ig_{\mu\nu}}{(p_{e}+p_{\overline{e}})^{2}
-M_{v_{i}}^{2}}\overline{v}_{e}
(p_{\overline{e}})\Lambda_{v_{i}e\overline{e}}^{\nu}u_{e}(p_{e});\\
M_{c}^{v_{i}T}&=&\overline{u}_{t}(p_{t})\Lambda_{v_{i}
t\overline{T}}^{\mu}\frac{i[(p_{\overline{t}}+p_{h})\cdot \gamma
+M_{T}]}{(p_{\overline{t}}+p_{h})^{2}
-M_{T}^{2}}\Lambda_{hT\overline{t}}v_{t}
(p_{\overline{t}})\nonumber\\
&&\hspace{1cm}\frac{-ig_{\mu\nu}}{(p_{e}+p_{\overline{e}})^{2}
-M_{v_{i}}^{2}}\overline{v}_{e}(p_{\overline{e}})
\Lambda_{v_{i}e\overline{e}}^{\nu}u_{e}(p_{e});\\
M_{d}^{v_{i}T}&=&\overline{u}_{t}(p_{t})\Lambda_{ht\overline{T}}\frac{i[(p_{t}+p_{h})\cdot
\gamma +M_{T}]}{(p_{t}+p_{h})^{2}
-M_{T}^{2}}\Lambda_{v_{i}T\overline{t}}^{\mu}v_{t}(p_{\overline{t}})
\nonumber\\
&&\hspace{1cm}\frac{-ig_{\mu\nu}}{(p_{e}+p_{\overline{e}})^{2}
-M_{v_{i}}^{2}}\overline{v}_{e}
(p_{\overline{e}})\Lambda_{v_{i}e\overline{e}}^{\nu}u_{e}(p_{e});\\
M_{e}^{v_{i}Z}&=&\overline{u}_{t}(p_{t})\Lambda_{Zt\overline{t}}^{\mu}v_{t}
(p_{\overline{t}})\frac{-ig_{\mu\mu'}}{(p_{t}+p_{\overline{t}})^{2}-M_{Z}^{2}
}\Lambda_{Zv_{i}h}^{\mu'\nu'}\nonumber\\
&&\hspace{1cm}
\frac{-ig_{\nu\nu'}}{(p_{e}+p_{\overline{e}})^{2}-M_{v_{i}}^{2}}\overline{v}_{e}
(p_{\overline{e}})\Lambda_{v_{i}
e\overline{e}}^{\nu}u_{e}(p_{e});\\
M_{f}^{v_{i}Z_{H}}&=&\overline{u}_{t}(p_{t})\Lambda_{Z_{H}t
\overline{t}}^{\mu}v_{t}(p_{\overline{t}})\frac{-ig_{\mu\mu'}}
{(p_{\overline{t}}+p_{t})^{2}-M_{Z_{H}}^{2}}
\Lambda_{Z_{H}v_{i}H}^{\mu'\nu'}\nonumber\\
&&\hspace{1cm}\frac{-ig_{\nu\nu'}}{(p_{e}+p_{\overline{e}})^{2}-
M_{v_{i}}^{2}}\overline{v}_{e}
(p_{\overline{e}})\Lambda_{v_{i}e\overline{e}}^{\nu}u_{e}(p_{e});\\
M_{g}^{v_{i}B_{H}}&=&\overline{u}_{t}(p_{t})\Lambda_{B_{H}t
\overline{t}}^{\mu}v_{t}(p_{\overline{t}})\frac{-ig_{\mu\mu'}}
{(p_{\overline{t}}+p_{t})^{2}-M_{B_{H}}^{2}}\Lambda_{B_{H}v_{i}H}^{\mu'\nu'}\nonumber\\
&&\hspace{1cm}\frac{-ig_{\nu\nu'}}{(p_{e}+p_{\overline{e}})^{2}
-M_{v_{i}}^{2}}\overline{v}_{e}
(p_{\overline{e}})\Lambda_{v_{i}e\overline{e}}^{\nu}u_{e}(p_{e}).
\end{eqnarray}
Where $\Lambda_{ijk}$ are the relevant coupling vertices, which
have been given in Ref.[12]. The LH model can generate correction
terms to the tree-level SM coupling vertices $\Lambda_{Zf\bar{f}}$
and $\Lambda_{ht\bar{t}}$, which can also produce corrections to
the process $e^{+}e^{-}\rightarrow t\bar{t}h$. In our numerical
calculation, we will take into account this correction effects.

From above equations, we can see that the $t\overline{t}h$
production cross section $\sigma$ involves four of the free
parameters of the $LH$ model, except the $SM$ input parameters
$\alpha_{e}$, $S_{W}$, $M_{Z}$, $m_{t}$, and $M_{h}$. They are the
vacuum condensate scale parameter $f$, the mixing parameters $c'$
and $c$ between the charged and neutral vector bosons, and the
mixing parameter
$x_{L}=\lambda_{1}^{2}/(\lambda_{1}^{2}+\lambda_{2}^{2})$ between
the $SM$ top quark and the heavy vector-like quark $T$.
$\lambda_{1}$ and $\lambda_{2}$ are the Yukawa coupling
parameters. At the order of $\nu^{2}/f^{2}$, the $T$ quark mass
$M_{T}$, the $B_{H}$ mass $M_{B_{H}}$, and the $Z_{H}$ mass
$M_{Z_{H}}$ mainly depend on the free parameters $f$, $x_{L}$;
$f$, $c'$; and $f$, $c$, respectively. The mixing parameters $c$
and $c'$ also control the couplings of the new heavy gauge bosons
$Z_{H}$ and $B_{H}$ to other particles.

In the $LH$ model, the custodial $SU(2)$ global symmetry is
explicitly broken, which can generate large contributions to the
electroweak observables. If one assumes that the $SM$ fermions are
charged only under $U(1)_{1}$, then global fits to the electroweak
precision data produce rather severe constraints on the parameter
space of the $LH$ model[12,14,15]. However, if the $SM$ fermions
are charged under $U(1)_{1}\times U(1)_{2}$, the constraints
become relaxed. The scale parameter $f=1\sim 2TeV$ is allowed for
the mixing parameters $c$, $c'$, and $x_{L}$ in the ranges of
$0\sim 0.5$, $0.62\sim 0.73$, and $0.3\sim0.6$, respectively[16].
Taking into account the constraints on the free parameters $f$,
$c$, $c'$ and $x_{L}$, we will give our numerical results in the
following section.

\noindent{\bf III Numerical results and summary}

To obtain numerical results, we need to specify the relevant $SM$
input parameters.  These parameters are $m_{t}=178GeV$,
$\alpha_{e}=1/128.8$, $S_{W}^{2}=0.2315$, $M_{Z}=91.187GeV$[17].
Considering the experimental constraints on the Higgs boson mass
$M_{h}$[4], we take $M_{h}=120GeV$. The c.m. energy $\sqrt{S}$ of
the $ILC$ experiment is assumed as $\sqrt{S}=800GeV$. In our
numerical estimations, we will take the parameters  $f$, $c$, $c'$
and $x_{L}$ as free parameters.

The relative correction $\delta\sigma/\sigma^{SM}$ is plotted in
Fig.2 as a function of the mixing parameter $c$ for $f=1TeV$ and
three values of the mixing parameter $c'$, in which
$\delta\sigma=\sigma^{tot}-\sigma^{SM}$ and $\sigma^{SM}$ is the
tree-level production cross section of $t\overline{t}h$ predicted
by the $SM$. We have taken the mixing parameter $x_{L} =0.3,\
0.4,\ 0.5,\ 0.6$ in Fig.2(a), Fig.2(b), Fig.2(c) and Fig.2(d),
respectively. From Fig.2 one can see that the total contribution
of the new particles predicted by the $LH$ model can enhance or
suppress the production cross section of the process
$e^{+}e^{-}\rightarrow t\overline{t}h$, which mainly depends on
the values of the mixing parameters $c$ and $c'$ for the fixed of
the parameter $ f $. The absolute value of the relative correction
 $\delta\sigma/\sigma^{SM}$ is not sensitive to the
mixing parameter $x_{L}$, while is sensitive to the mixing
parameters $c$ and $c'$. For $c\geq0.35$, the value of
$\delta\sigma/\sigma^{SM}$ quickly increases as $c$ increasing.
This is because, for the fixed parameters $c'$, $x_{L}$ and $f$,
the correction cross section $\delta \sigma$ is proportional to
the factors $c^{4}$ and $c^{2}(c^{2}-s^{2})$ at the order of
$\nu^{2}/f^{2}$. The factor $c^{2}(c^{2}-s^{2})$ goes to extremum
when the mixing parameter $c$ gets close to $0.5$. For $c<0.35$,
in most of the parameter space of the $LH$ model, the absolute
value of the relative correction
 $\delta\sigma/\sigma^{SM}$ is smaller than $5\%$, which is very difficult to be
detected in the future $ILC$ experiments. Certainly, it is very
easy to observe the correction effects of the $LH$ model on the
process $e^{+}e^{-}\rightarrow t\overline{t}h$ for $c\geq 0.35$.

\begin{figure}[htb]
\vspace{-0.5cm}
  \centering
   \includegraphics[width=3.3in]{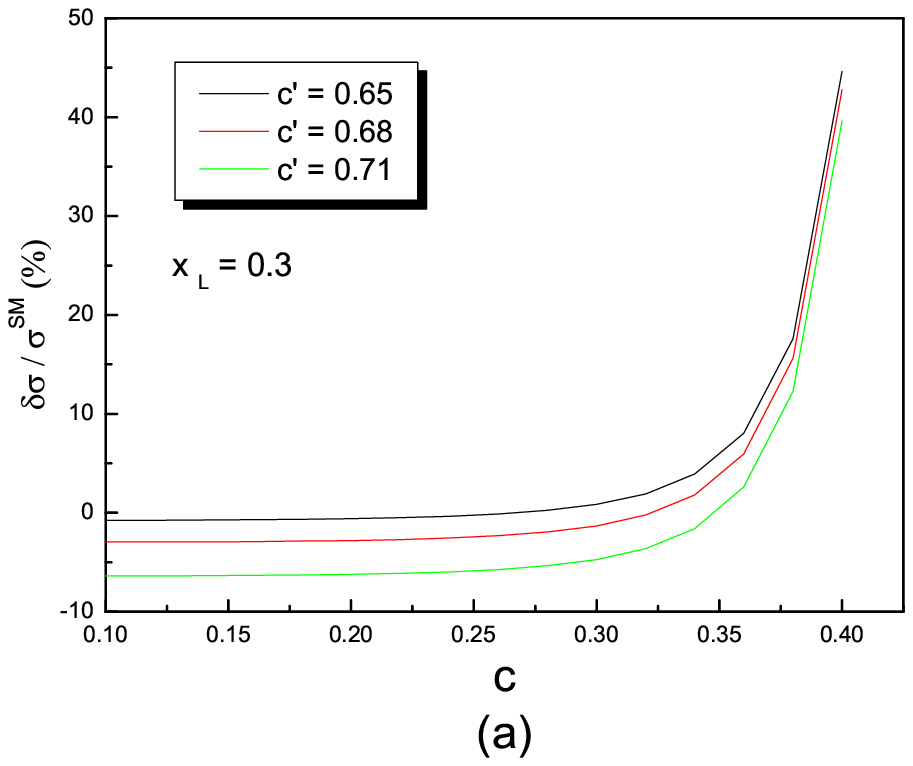}
    \hspace{-0.4in}
   \includegraphics[width=3.3in]{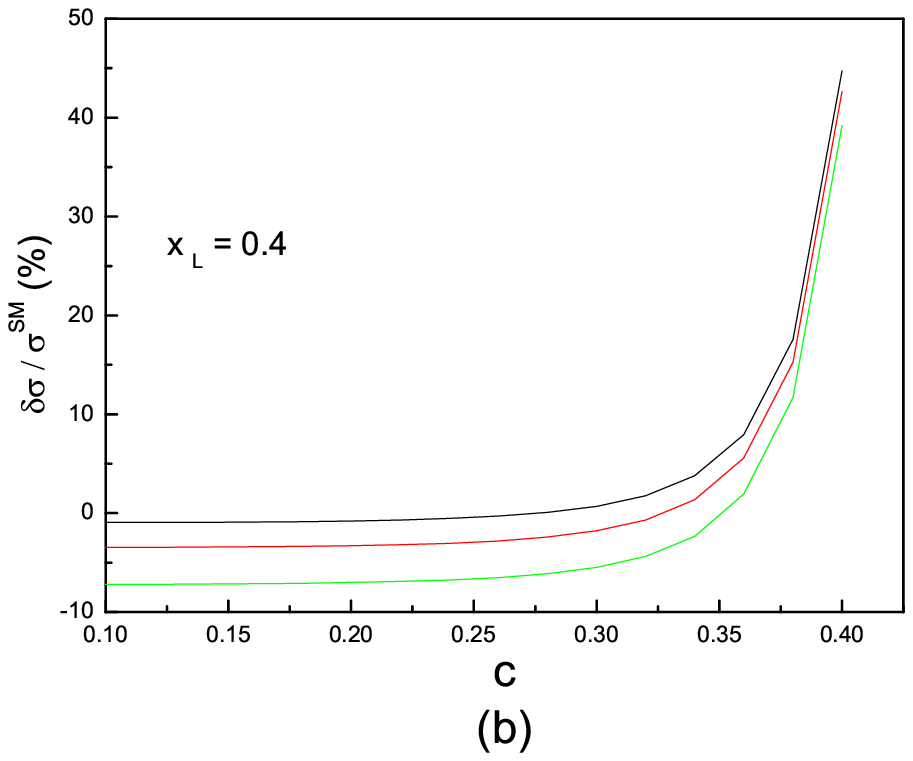}
\end{figure}

\begin{figure}[htb]
\vspace{-1.5cm}
  \centering
   \includegraphics[width=3.3in]{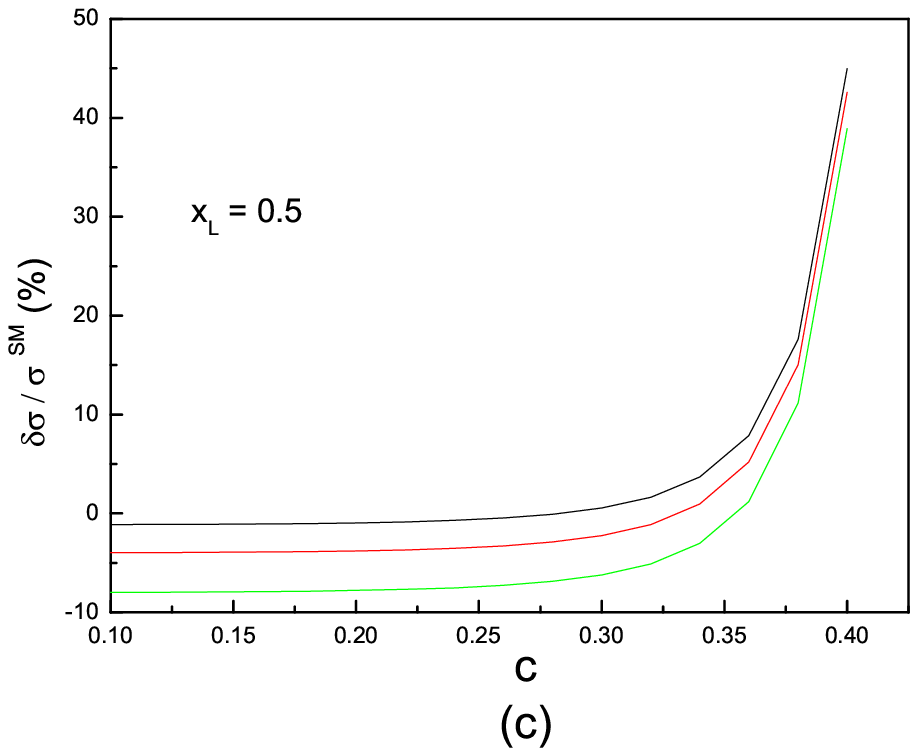}
    \hspace{-0.6in}
   \includegraphics[width=3.3in]{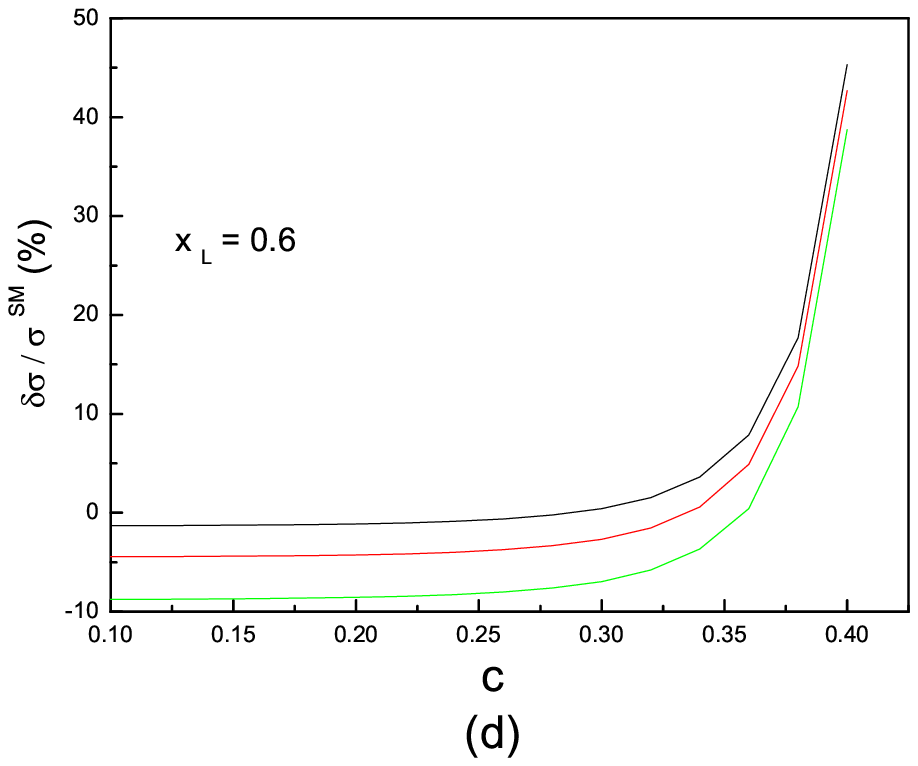}\vspace{-1cm}
\hspace{0.5cm}\caption{The relative correction
$\delta\sigma/\sigma^{SM}$ as a
  function of the mixing parameter $c$ for \hspace*{1.8cm}$f=1TeV$ and
  different values of  the mixing parameters $c'$ and $x_{L}$.}
\end{figure}

\begin{figure}[htb]
\begin{center}
\epsfig{file=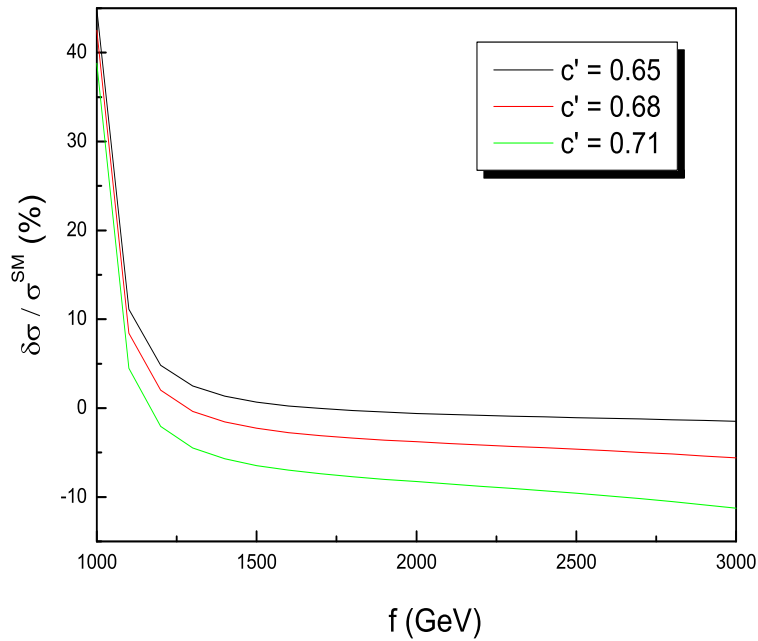,width=320pt,height=280pt} \vspace{-0.5cm}
\hspace{0.5cm} \caption{The relative correction parameter
$\delta\sigma/\sigma^{SM}$ as a function of the scale parameter
\hspace*{1.8cm}$f$ for $x_{L}=0.5,\ c=0.4$ and three values of the
mixing parameter $c'$.} \label{ee}
\end{center}
\end{figure}

In general, the contributions of the $LH$ model to observables are
dependent on the factor $1/f^{2}$. To see the effect of varying
the scale parameter $f$ on the relative correction
$\delta\sigma/\sigma^{SM}$, we plot $\delta\sigma/\sigma^{SM}$ as
a function of $f$ for three value of the mixing parameter $c'$ in
Fig.3. Taking into account the conclusions obtained from Fig.2, we
have taken $x_{L}=0.5$ and $c=0.4$ in Fig.3. One can see from
Fig.3 that the total contribution of the $LH$ model to the
$t\overline{t}h$ production cross section decreases as the scale
parameter $f$ increasing for $f\leq 1.2TeV$ with $c'=0.68$.
However, for $1.2TeV< f \leq 4.8 TeV$, the value of
$\delta\sigma/\sigma^{SM}$ is negative and its absolute value
increases as $f$ increasing. This is because the contributions of
the LH model to the $t\bar{t}h$ production cross section mainly
come from $Z_{H}$ exchange and $B_{H}$ exchange. $Z_{H}$ exchange
has positive contributions which quickly decrease as the scalar
parameter $f$ increasing, while $B_{H}$ exchange has negative
contributions which slowly decrease as $f$ increasing. For
$f>5TeV$, the total contributions to the $t\bar{t}h$ production
cross section get close to zero. Thus, in general, the total
contributions decouple for large of the scale parameter $f$, which
is similar to the contributions of the Lh model to for other
observables.

The $LH$ model[1] is one of the simplest of the little Higgs
models, which predicts  the existence of the four new heavy gauge
bosons $Z_{H}$, $B_{H}$, and $W^{\pm}_{H}$, a vector-like top
quark and a triplet of heavy scalars except the $SM$ particles.
Some of these new particles can generate significant corrections
to the electroweak precision observables and thus the precision
measurement data can give severe constraints on the parameter
space of this type of models. In the parameter space of the $LH$
model($f=1\sim 2TeV,\ c=0\sim 0.5,\ c'=0.62\sim 0.73$) preferred
by the electroweak precision data, we study the process
$e^{+}e^{-}\rightarrow t\overline{t}h$. We find that the
contributions of the $LH$ model to this process mainly come from
Fig.1$(a)$ and $(b)$ generated by $Z_{H}$ exchange and $B_{H}$
exchange, which can enhance or suppress the $t\overline{t}h$
production cross section $\sigma^{SM}$ predicted by the $SM$ at
the tree-level. In sizable regions of the parameter space, the
absolute value of the relative correction
$\delta\sigma/\sigma^{SM}$ is larger than $5\%$, which might be
detected in the future $ILC$ experiments.

Certainly, the modification to the relation between the SM free
parameters can also produce contributions to the processes
$e^{+}e^{-}\rightarrow t\overline{t}h$. However, our calculation
results show that the contributions are smaller than those of
$Z_{H}$ exchange and $B_{H}$ exchange at least by one order of
magnitude in wide range of the parameter space of the LH model.
Thus, comparing with the direct contributions of $Z_{H}$ and
$B_{H}$, the contributions from the modification to the relation
between the SM free parameters can be safely neglected.

\vspace{0.5cm} \noindent{\bf Acknowledgments}

This work was supported in part by Program for New Century
Excellent Talents in University(NCET), the National Natural
Science Foundation of China under the grant No.90203005 and
No.10475037, and the Natural Science Foundation of the Liaoning
Scientific Committee(20032101).

\null
\end{document}